\documentclass[aps,prb,floats,onecolumn,preprint]{revtex4}

\usepackage{amsmath}
\usepackage{epsfig}

\usepackage[colorlinks=true,hypertex,hyperref,dvips,ps2pdf]{hyperref}
\begin{document}

Appeared in the Journal of Chemical Physics, 125, 194508

\title{Time dependent diffusion in a disordered medium with partially
    absorbing walls: A perturbative approach}
\author{ Jiang Qian$^{1,2}$ and Pabitra N. Sen$^{2}$,}
\affiliation{
  $^1$Lyman Laboratory of Physics, Harvard University, Cambridge, MA
    02138, USA   \\
    $^2$Schlumberger-Doll Research, Ridgefield, CT 06877-4108, USA
}
\begin{abstract}
We present an analytical study of the time dependent diffusion
coefficient in a dilute suspension of spheres with partially
absorbing boundary condition. Following Kirkpatrick~(J. Chem. Phys. 76, 
4255)~we obtain a perturbative expansion for the time dependent particle 
density using volume fraction $f$ of spheres as an expansion parameter.  
The exact single particle $t$-operator for partially
absorbing boundary condition is used to obtain a closed form
time-dependent diffusion coefficient $D(t)$ accurate to first
order in the volume fraction $f$. Short and long time limits of $D(t)$
are checked against the known short-time results for partially or
fully absorbing boundary conditions  and long-time results for
reflecting boundary conditions. For fully absorbing boundary
condition the long time diffusion coefficient is found to be
$D(t)=5 a^2/(12 f D_{0} t) +O((D_0t/a^2)^{-2})$, to the first order of
perturbation theory. Here $f$ is small but non-zero, $D_0$ the
diffusion coefficient in the absence of spheres, and $a$ the
radius of the spheres. The validity of this perturbative result is
discussed.
\end{abstract}
\date{\today}
\maketitle
\section{Introduction}
\label{sec:intro}

There are numerous processes in biology, physics, geophysics,
chemical engineering and materials science where a diffusing
species reacts upon contact with the surface of another phase
(typically a solid) and the effects of simultaneous diffusion and
reaction are important. Studying the time-dependence of
concentration of the species reacting with a porous host via
diffusion, which goes back to Smoluchowsky
(1913)\cite{smoluchowsky}, remains an important problem both for
practical and theoretical reasons: for reviews, see Weiss
\cite{weiss}, Kayser and Hubbard\cite{kayser}, Fixman
\cite{fixman}, Torquato\cite{kansal_torquato},
deGennes\cite{degennes}. Heterogeneous catalysis, transport and
absorption of nutrients and drugs are well known examples. Another
example is relaxation of Nuclear magnetization of interstitial fluid by 
interaction with walls in the NMR experiments. The connection
between surface relaxation and NMR goes back to the seminal 1951
paper ``On nuclear relaxation in gases by surface catalysis" by
Bloch\cite{bloch} himself. NMR Relaxation is now routinely used in
numerous settings ranging from clinical MRI
\cite{nmr_encyclopedia} to geological explorations for energy
\cite{kleinberg}. The purpose of this paper is to investigate 
time-dependence of the diffusion coefficient of molecules confined in
a well-connected porous host material with reactive walls.

Diffusion measurements on surviving number of random walkers can give 
deep insight
into this complex process of diffusion-relaxation.  Experimentally
observing the dynamics (diffusion) of molecules reveals a distinct
set of information that cannot be obtained from the traditional
relaxation (i.e. number of surviving walkers),
measurements\cite{concepts}. The complex porous systems, in whose
interstices the fluid reside, have in general, convoluted and
extensive well connected pores. The time dependence of the rate of
relaxation in a well connected geometry which does not have
arbitrarily large voids (see below) is not fully understood, not
to mention diffusion. The relaxation processes are well understood
for isolated cells or pores (isolated cavities), where an
eigen-mode decomposition
\cite{brownstein79,zielinskyjcp2002,luca} of the
diffusion equation with partially absorbing boundary condition is
possible. Disentangling the complexity due to the combined effects
of disorder, diffusion and decay remains a major hurdle in
connected porous media\cite{concepts}.

A general theory of time-dependent diffusion and relaxation in
porous media does not exist. At short-times, there are asymptotic
results that are robust for all boundary conditions and arbitrary
geometry \cite{concepts,mitra1,mitra2}. For
long-time, however, no general technique exists, except for simple
models of a dilute periodic array of spheres  with weak
absorption\cite{sen_halperin}. In random systems, a dilute
suspension of randomly placed spheres is most well studied. For
reflecting (non-absorbing) boundary condition de Swiet and
Sen\cite{deswiet96}  worked out a perturbation expansion based on
Bixon and Zwanzig\cite{BZ} that gives  short and long time
behavior of the time dependent diffusion coefficient.

The main goal  of this paper is to extend these
results\cite{deswiet96} for partially absorbing boundary
conditions. For partially absorbing boundary conditions perturbation 
expansions fail at long
times\cite{kirk,sen_halperin}. The long time behavior of diffusion
in such a system has not been previously addressed either
analytically or numerically. Decay in particle number density with
fully absorbing boundary condition has been studied extensively, see for 
example the review by Weiss\cite{weiss}. The simplest
effective medium approach of Smoluchowsky gives an exponential
decay for the total number of surviving particles $N(t)$. Bixon
and Zwanzig pioneered the perturbative approach which gave a
surprisingly slow power law decay. This discrepancy has triggered
an intense interest in the problem. Kirkpatrick\cite{kirk}
perfected the perturbation theory, and he showed that by summing a
select group of terms (the most divergent) this power law tail was
removed and the exponential decay was recovered. However, the
higher order terms in the perturbation series diverge as well and are
difficult to re-sum  and  one cannot  rule out long-time power law
tails \cite{kirk}.

Later, allowing arbitrarily large fluctuations in local density,
Grassberger and Proccacia, Kayser and Hubbard, and others
\cite{grassberger82,kayser,varadhan83} conclusively
showed that the long time limiting behavior of $N(t)$ in such
system shows a non-perturbative behavior: in a $d$-dimension
system $N(t)$ shows a stretched exponential decay
$N(t)\to\exp[-t^{d/(d+2)}]$ as $t\to\infty$.\footnote{This result
also holds for partially absorbing boundary
condition\cite{kayser}} This result cannot be captured by any
(known) re-summation of the perturbation expansion. However  three
issues are important--- (i) real systems do not have large voids
and this stretched exponential result is of limited use (ii)
Fixman\cite{fixman} has argued that the crossover from the
effective medium theory (simple exponential decay in time) to the
exact limiting behavior (i.e. the stretched exponential) can be
extraordinarily slow, i.e. the breakdown occurs at so long a
time that $N(t)$ is too small to be of any experimental relevance.
(iii) we are interested in $D(t)$ and not, \textit{per se}, in
$N(t)$. $D(t)$ is given by mean-square displacement divided by
$N(t)$, there is no known rigorous bound for this ratio. It is
possible that perturbative expansion can give a reasonable answer
for $D(t)$. We note in passing that Grassberger and
Proccacia\cite{grassberger82} give a heuristic argument that
$D(t)$ approaches zero as $t^{-d/(d+2)}$ at long times. Even such a
heuristic conjecture does not exist for partially absorbing systems
nor for systems that do not have large fluctuations in density.

Therefore, we may argue that, for experimentally accessible times
$t$, the perturbation expansion describes diffusion reasonably
well and this is what we proceed to carry out. In the next section
\ref{sec:analytical} we carry out a perturbation expansion for a
dilute suspension of randomly placed spheres. We obtain
an analytical expression of the approximate Green's function from
first order terms of that expansion. From the Green's function we
find the simple short and long time limiting behavior of time
dependent diffusion coefficient $D(t)$.

\bigskip

\section{Analytical results for suspension of sphere}
\label{sec:analytical}

In this section we study the problem  of diffusion in dilute
system with the partially absorbing boundary condition analytically. 
Only dilute
systems are amenable to analytical methods. We will address the
range of validity of such methods at the end of this section.

Let us consider now diffusion of particles around a randomly distributed 
dilute suspension of $N$ spheres, all of which of a given radius $a$, 
centered at positions
$\vec{R}_{i}$, $1\le i\le N$.
The density $C(\vec{r},t)$
of particles is governed by diffusion equation
\begin{equation}
\label{eqn:diffusion}
\frac{\partial C(\vec{r},t)}{\partial
t}=D_{0}\nabla^{2}C(\vec{r},t)
\end{equation}
and the partially absorbing boundary condition is imposed:
\begin{equation}
\label{eqn:boundary_original}
D_{0}\vec{n}_{i}\cdot\nabla C(|\vec{r}-\vec{R}_{i}|=a,t)=\rho
C(|\vec{r}-\vec{R_{i}}|=a,t),
\end{equation}
\begin{equation}
C(|\vec{r}-\vec{R_{i}}|<a,t)=0,
\end{equation}
for $i$ from 1 to $N$, where $\vec{n}_{i}$ is the unit vector normal to the
$i$th sphere, pointing outward. The initial condition is
\begin{equation}
\label{eqn:initial}
C(\vec{r},t=0)=C_{0}(\vec{r}).
\end{equation}
Following Kirkpatrick\cite{kirk} we go into the Laplace domain for time 
and the Fourier domain for space and express density in terms of 
$t$-operator of
individual sphere as scattering center,
\begin{equation}
\label{eqn:scatter1}
C(\vec{q},\epsilon)=G_{0}(q)C_{0}(\vec{q})+\sum_{i=1}^{N}G_{0}(q)T_{i}(\vec{q})C_{i}(\vec{q},\epsilon),
\end{equation}
\begin{equation}
\label{eqn:scatter2}
C_{i}(\vec{q},\epsilon)=G_0(q)C_{0}(\vec{q})+\sum_{j\ne
i}^{N}G_{0}(q)T_{j}(\vec{q})C_{j}(\vec{q},\epsilon).
\end{equation}
Here $\epsilon$ is the reciprocal Laplace variable for time and
$\vec{q}$ is the reciprocal Fourier variable for space and
$G_{0}(q)=(\epsilon +Dq^{2})^{-1}$ is the free particle Green's
function, i.e. diffusion in unbounded space, and $T_{i}$ is the
Fourier space $t$-operator of a single sphere centered at $R_{i}$.
Combining Eq.~(\ref{eqn:scatter1}) and Eq.~(\ref{eqn:scatter2}) we
have the following perturbative expansion for the density\cite{kirk}:
\begin{widetext}
\begin{equation}
\label{eqn:expansion}
\begin{split}
C(\vec{q},\epsilon)=\Bigg\{G_{0}(q)+\sum_{i=1}^{N}G_{0}(q)T_{i}(\vec{q})G_{0}(q)+
\sum_{i=1}^{N}\sum_{j\ne i}^{N}
G_{0}T_{i}(\vec{q})G_{0}(q)T_{j}(\vec{q})G_{0}(q)+\\
\sum_{i=1}^{N}\sum_{j\ne i}^{N}\sum_{k\ne j}^{N}
G_{0}T_{i}(\vec{q})G_{0}(q)T_{j}(\vec{q})G_{0}(q)T_{k}(\vec{q})G_{0}(q)+\cdots
\Bigg\}C_{0}(\vec{q}),
\end{split}
\end{equation}
\end{widetext}
This equation has the same form of the binary collision expansion in
kinetic theory of gases\cite{kirk}. This naive expansion encounters
a problem at long time\cite{kirk,ernst} and it is better to sum
the series to avoid such difficulties\cite{BZ,kirk}. The averaged 
results reads:
\begin{equation}
\label{eqn:pert} C(\vec{q},\epsilon)=\Big\{\epsilon +
D_{0}q^{2}-\sum_{i=1}^{\infty}n^{i}\int
\prod_{j=1}^{i} 
d\vec{R}_{i}~B_{i}(\vec{q},\vec{R_1}\ldots\vec{R}_{i},\epsilon)\Big\}^{-1}C_{0}(q).
\end{equation}
Here $n=N/V$ is the number of spheres per unit
volume and $B_{i}(\vec{q},\vec{R}_{i},\epsilon)$ represents the ``self
energy" operator for multiple scattering and includes correlation
effects of averaging of $i$ scattering center. For example, when
Eq. (\ref{eqn:pert})is expanded, the operator form gives
\begin{equation}
\label{eqn:pert_1st}
\int d\vec{R}_{1}B_1(\vec{q},\vec{R}_{1},\epsilon)
=\int d\vec{R}_{1} T_{1}(\vec{q}),
\end{equation}

\begin{widetext}
\begin{equation}
\label{eqn:pert_2nd}
\begin{split}
\int 
d\vec{R}_{1}d\vec{R}_{2}B_2(\vec{q},\vec{R}_{1},\vec{R}_{2},\epsilon)=
\int d\vec{R}_{1} \int d\vec{R}_{2}\{
T_{1}(\vec{q})G_{0}(q)T_{2}(\vec{q})G_{0}(q)T_{1}(\vec{q})+\\
T_{1}(\vec{q})G_{0}(q)T_{2}(\vec{q})G_{0}(q)T_{1}(\vec{q})G_{0}(q)T_{2}(\vec{q})+
\cdots\}.
\end{split}
\end{equation}
\end{widetext}
As usual, here $T_{i}$ is the $t$-operator of a single sphere at
position $\vec{R}_{i}$ and the integrals are averaging over trap
positions. In this paper we confine ourself to the effect of the
first term of the expansion, i.e. we include multiple scattering
from a single sphere to all orders, but do not consider
correlation effects of multiple spheres.

As required by translational invariance, the dependence of the 
$t$-operator on
the position of sphere $\vec{R}_{i}$ is very simple:
\begin{equation}
\label{eqn:t_translation}
t_{i}(\vec{q},\vec{q}\prime,\epsilon,\vec{R}_{i})=exp[-i\vec{R}_{i}\cdot
(\vec{q}-\vec{q}\prime)]t(\vec{q},\vec{q}\prime,\epsilon).
\end{equation}
Therefore the lowest order position averaged $t$-operator simply gives
a factor of $(2\pi)^{3}\delta^{3}(\vec{q}-\vec{q}\prime)$ and we only need the
diagonal element of the $t$-operator $\langle t 
\rangle=t(q,q,\epsilon)$.

Transformed into Laplace space the diffusion equation reads:

\begin{equation}
\label{eqn:laplace_diffusion}
(D_{0}\nabla^2-\epsilon)C(\vec{r},\epsilon)=-C_{0}(\vec{r}),
\end{equation}

The boundary condition in Eq.~(\ref{eqn:boundary_original}) becomes, in
Laplace domain:

\begin{equation}
\label{eqn:boundary_new} \lambda a \vec{n}_{i}\cdot\nabla
C(\vec{r},\epsilon)-C(\vec{r},\epsilon)=0 ; \;\;\;
\lambda=\frac{D_{0}}{\rho a}
\end{equation}
on the surface $|\vec{r}-\vec{R}_{i}|=a$. Here we introduced a
dimensionless parameter $\lambda= D_{0}/\rho a $ as a measure of
strength of absorption on surface. $\lambda=0$ corresponds to
totally absorbing boundary condition and $\lambda=\infty$
corresponds to reflecting boundary condition.

It is convenient to choose the initial condition
$C_{0}(\vec{r})=\delta^{3}(\vec{r}-\vec{r}\prime)$, which when Fourier
transformed against $\vec{r}\prime)$ reads:
\begin{equation}
\label{eqn:laplace_initial}
C_{0}(\vec{r},\vec{q}\prime,\epsilon)=\frac{e^{i\vec{q}\prime\cdot\vec{r}}}{(2\pi)^3}.
\end{equation}
A special solution for inhomogeneous equation (\ref{eqn:laplace_diffusion}) is
\begin{equation}
\label{eqn:special}
C_{special}(\vec{r},\vec{q}\prime,\epsilon)=
\frac{1}{(2\pi)^{3}}\frac{e^{i\vec{q}\prime\cdot\vec{r}}}{\epsilon+D_{0}{q\prime}^{2}}.
\end{equation}
and the general solution of the homogeneous equation satisfying boundary 
condition
that density to be finite at infinity is:
\begin{equation}
\label{eqn:general}
C_{general}(\vec{r},\epsilon)=\sum_{i=0}^{\infty}b_{i}P_{i}(cos\theta^{\prime})\kappa_{i}(\frac{|\vec{r}-\vec{R}_{i}|\sqrt{s}}{a}),
\end{equation}
where $\kappa_{i}(x)$ is the $i$th order modified spherical Bessel 
function and
$P_{i}(x)$ is the $i$th order Legendre polynomial. $\theta^{\prime}$ is the
angle between $\vec{q}^{\prime}$ and $\vec{r}-\vec{R}_{i}$ and we introduce
dimensionless parameter $s=\epsilon a^2/D_{0}$ to replace $\epsilon$.

Choosing constants $b_{i}$ of Eq.~(\ref{eqn:general}) to make a linear
combination with Eq.~(\ref{eqn:special}) that satisfy boundary condition
Eq.~(\ref{eqn:boundary_new}) we have for $|\vec{r}-\vec{R}_{i}|>a$:

\begin{widetext}
\begin{equation}
\label{eqn:density}
C(\vec{r},\vec{q}\prime,\epsilon)=
\frac{1}{(2\pi)^{3}}\frac{e^{i\vec{q}\prime\cdot\vec{R}_{i}}}{\epsilon+D_{0}{q\prime}^{2}}
\Bigg[
e^{i\vec{q}\prime\cdot(\vec{r}-\vec{R}_{i})}-
\sum_{n=0}^{\infty} (2n+1)i^{n}P_{n}(cos\theta^{\prime})
\frac{\lambda(q'a)j'_{n}(q'a)-j_{n}(q'a)}
{\lambda \sqrt{s}\kappa'_{n}(\sqrt{s})-\kappa_{n}(\sqrt{s})}
\kappa_{n}(\frac{|\vec{r}-\vec{R}_{i}|\sqrt{s}}{a})
\Bigg],
\end{equation}
\end{widetext}
where $j_{n}(x)$ is the $n$th order spherical Bessel function.

Fourier transforming Eq.~(\ref{eqn:density}) with respect to $r$
we can easily obtain the single particle $t$ operator:
\begin{widetext}
\begin{equation}
\begin{split}
\label{eqn:t_original} t(\vec{q},\vec{q}\prime,\epsilon)=
-\frac{4\pi
aD_{0}e^{i(\vec{q}-\vec{q}\prime)\cdot\vec{R}_{i}}}{(2\pi)^{3}}
\Bigg\{
\frac{s+k^2}{|\vec{k}-\vec{k}\prime|}j_{1}(|\vec{k}-\vec{k}\prime|)+
\sum_{n=0}^{\infty}(2n+1)P_{n}(cos\theta)\times\\
\frac{\lambda k'j'_{n}(k')-j_{n}(k')} {\lambda
\sqrt{s}\kappa'_{n}(\sqrt{s})-\kappa_{n}(\sqrt{s})}
[\sqrt{s}j_{n}(k)\kappa_{n+1}(\sqrt{s})-kj_{n+1}(k)\kappa_{n}(\sqrt{s})]
\Bigg\};\;\; \vec{k}=\vec{q}a ; \; \vec{k}\prime=\vec{q}\prime a
\end{split}
\end{equation}
\end{widetext}
Here we made the obvious choice for dimensionless momentum
$\vec{k}=\vec{q}a$ and $\vec{k}\prime=\vec{q}\prime a$ and
$\theta$ is the angle between $\vec{k}$ and $\vec{k}\prime$. This
form for the $t$-operator manifestly obeys 
Eq.~(\ref{eqn:t_translation}).
Setting $\vec{k}=\vec{k}\prime$ we
easily obtain the diagonal element of the $t$-operator

\begin{widetext}
\begin{equation}
\begin{split}
\label{eqn:t_average}
\langle t \rangle=-(4\pi aD_{0})\Bigg\{\frac{1}{3}(s+k^2)+
\sum_{n=0}^{\infty}(2n+1)
\frac{\lambda k j'_{n}(k)-j_{n}(k)}
{\lambda \sqrt{s}\kappa'_{n}(\sqrt{s})-\kappa_{n}(\sqrt{s})}
[\sqrt{s}j_{n}(k)\kappa_{n+1}(\sqrt{s})-kj_{n+1}(k)\kappa_{n}(\sqrt{s})]
\Bigg\}
\end{split}
\end{equation}
\end{widetext}
\bigskip

To obtain the effective diffusion coefficient, we expand
Eq.~(\ref{eqn:t_average}) to order $k^2$ and obtain the following
leading order behavior:
\begin{widetext}
\begin{equation}
\label{eqn:t_expansion}
\langle t \rangle=-(4\pi aD_{0})\Bigg\{\frac{1}{3}(s+k^2)+
\frac{1+\sqrt{s}}{1+\lambda+\lambda \sqrt{s}}-
k^2\Big[
\frac{-1+(3+\sqrt{s})\lambda)+(1+2\sqrt{s}+s)\lambda^{2}}
{3(1+\lambda+\lambda\sqrt{s})(1+\sqrt{s}+(2+2\sqrt{s}+s)\lambda}
\Big]
\Bigg\}
\end{equation}
\end{widetext}
Taking the lowest order in Eq.~(\ref{eqn:pert}), combining with
Eq.~(\ref{eqn:pert_1st})
and definition of Green's function
we have
$\langle G \rangle=(G_{0}^{-1}-\frac{N}{V}\langle t\rangle)^{-1}$.  
Substituting from
Eq.~(\ref{eqn:t_expansion}) we have, again to the order of $k^2$:
\begin{widetext}
\begin{equation}
\begin{split}
\label{eqn:green}
\langle G\rangle=\frac{a^2}{D_{0}}\{s+f[s+\frac{3(1+\sqrt{s})}{(1+\lambda+
\lambda\sqrt{s})}]\}^{-1}
\Bigg\{
1-k^2\bigg[
\frac
{1+f(1-\frac{-1+(3+\sqrt{s})\lambda+(1+\sqrt{s})^2\lambda^2}
{(1+\lambda+\lambda\sqrt{s})(1+\sqrt{s}+\lambda(2+2\sqrt{s}+s))}
)}
{s+f(s+\frac{3(1+\sqrt{s})}{(1+\lambda+
\lambda\sqrt{s})})}
\bigg]
\Bigg\}.
\end{split}
\end{equation}
\end{widetext}
Here we define the volume fraction of the spheres $f=\frac{4\pi 
a^{3}N}{3V}$.
This generalizes Kirkpatrick's\cite{kirk} Eq.~(A5) to finite absorption.
To obtain $D(t)$ notice that for the system
with partially absorbing boundary condition:
\begin{equation}
\label{eqn:r_square}
\langle r^2 \rangle=-6a^2\frac{\lim\limits_{k \to 0}\partial
G(k,t)/\partial k^2}{\lim\limits_{k \to 0} G(k,t)}.
\end{equation}
The numerator is the inverse Laplace transform of the following expression:
\begin{equation}
\label{eqn:num}
\frac
{\big[1+f(1-\frac{-1+(3+\sqrt{s})\lambda+(1+\sqrt{s})^2\lambda^2}
{(1+\lambda+\lambda\sqrt{s})(1+\sqrt{s}+\lambda(2+2\sqrt{s}+s))}
)\big]}
{\big[s+f(s+\frac{3(1+\sqrt{s})}{(1+\lambda+
\lambda\sqrt{s})})\big]^2}.
\end{equation}
The denominator, that is the number of surviving particles 
$N(\tilde{t})$, is the inverse Laplace transform of $G(k=0)$:
\begin{equation}
\label{eqn:den}
\frac{1}{s+f[s+\frac{3(1+\sqrt{s})}{(1+\lambda+
\lambda\sqrt{s})}]}.
\end{equation}

Performing the inverse Laplace transform of the above
equation,  we obtain for the denominator, the total number of
surviving particles:
\begin{widetext}
\begin{equation}
\begin{split}
\label{eqn:N_full} N(\tilde{t})=\frac{1}{{(1+f)
   (r_{1}-r_{2}) (r_{1}-r_{3})
   (r_{2}-r_{3}) \lambda }}
\bigg[ e^{r_{1}^2 \tilde{t}} r_{1} (r_{2}-r_{3}) (r_{1}
\lambda+\lambda +1) \text{erfc}\left(-r_{1}
\sqrt{\tilde{t}}\right)+ c.p.
\bigg],
\end{split}
\end{equation}
\end{widetext}
here $\tilde{t}=D_0t/a^2 $,
$\text{erfc}(x)=1-\text{erf}(x)$ is the complementary error
function, $c.p.$ denotes cyclical permutation $r_1\to r_2$, $r_2\to r_3$
and $r_3\to r_1$, and $r_1,r_2,r_3$ are the roots of the cubic equation:
\begin{equation}
\label{eqn:root1}
(1+\lambda)fr^3+(1+f+\lambda f)r^2+3fr+3f=0.
\end{equation}
To check this with the known results we look at several limits
where the forms of $N(t)$ are simple.

First consider the short time limit $\tilde{t}\ll 1$, \textit{i.~e.},
$t\ll a^2/D_{0}$. There are two subcases.  If the boundary is
nearly reflecting, specifically $\sqrt s\gg 1/\lambda$, i.e.
$\rho\ll \sqrt{D_0/t}$, the results is particularly simple:
\begin{equation}
N(\tilde{t})=1-f-\frac{3 f \tilde{t}}{\lambda }+\frac{4 f
\tilde{t}^{3/2}}{\sqrt{\pi } \lambda
   ^2}+O\left(\tilde{t}^{5/2}\right).
\end{equation}
Here $1-f$ factor comes from the excluded volume of the spheres. The
first term agrees with equation
(A9) of Mitra \textit{et.~al.} \cite{mitra2}~, which is only to the 
first order in $f$.

In the other extreme, when $\rho\gg \sqrt{D_0/t}$, that is the nearly 
fully
absorbing boundary condition, the short time total number of particle
becomes insensitive to $\rho$:
\begin{equation}\label{eqn:n}
N(\tilde{t})=1-f-\frac{(6f-12f^2)\sqrt{\tilde{t}}}{\sqrt{\pi}}+(-3f+15f^2)\tilde{t}+O(\tilde{t}^{3/2}),
\end{equation}
The first term in turns agrees with equation (A8) of Mitra
\textit{et.~al.}\cite{mitra2}, again to the first order in
$f$.

Next consider the long-time limits. The total particle
number decays, in the long time limit, $\tilde{t}\gg 1$, i.e.,
$t\gg a^2/D_0$, for fully absorbing boundary condition:
\begin{equation}
\label{eqn:total}
N(\tilde{t})=\frac{1}{6f\sqrt{\pi}}~\tilde{t}^{-3/2}+\frac{2-f+2(1+f)\lambda}{12f^2\sqrt{\pi}}~\tilde{t}^{-5/2}+O(\tilde{t}^{-7/2})
\end{equation}
The first term is twice the perturbative result in Bixon and
Zwanzig\cite{BZ}. Recall that the exact non-perturbative result of
Grassberger and Procaccia\cite{grassberger82} gives a stretched
exponential decay. Also note the $1/f$ dependence, for small but
finite $f$.

Finally consider the diffusion coefficient with partially
absorbing boundary conditions. Results from the inverse Laplace
transform of the numerator in Eq.~(\ref{eqn:num}) is too complex
to be reproduced fully here, so we summarize the result in a
simpler form:
\begin{align}\label{numerator}
&\lim\limits_{k \to 0}\frac{\partial G(k,\tilde{t})}{\partial
k^2}=
\sum_{i=1}^3\sum_{j=1}^2~A_{ij}I_{j}(r_{i},\tilde{t})+\sum_{i=4}^5~A_{i1}I_{1}(r_{i},\tilde{t});
\end{align}
In this equation, $r_{1-3}$ are the same three roots of the cubic
equation Eq.~(\ref{eqn:root1}) and $r_{4,5}=(-1-2\lambda \pm
\sqrt{1-4\lambda^2})/2\lambda$ are the two roots of the quadratic 
equation.  And the
constants $A_{ij}$, which depends on $r_{1}$ through $r_{5}$ above
as well as $\lambda$ and $f$, are the coefficients of partial
fractions of Eq.~(\ref{eqn:num}):
\begin{align}
&\frac
{\big[1+f(1-\frac{-1+(3+\sqrt{s})\lambda+(1+\sqrt{s})^2\lambda^2}
{(1+\lambda+\lambda\sqrt{s})(1+\sqrt{s}+\lambda(2+2\sqrt{s}+s))}
)\big]}
{\big[s+f(s+\frac{3(1+\sqrt{s})}{(1+\lambda+
\lambda\sqrt{s})})\big]^2}\\
&
=\sum_{i=1}^5~\frac{A_{i1}}{\sqrt{s}-r_{i}}+
\sum_{i=1}^3~\frac{A_{i2}}{(\sqrt{s}-r_{i})^2} .
\end{align}
$I_{1,2}$ are given by the following expression obtained by inverse
Laplace transform of the partial fractions:
\begin{align}
I_1(r_i,\tilde{t})&=\frac{1}{\sqrt{\pi
\tilde{t}}}+r_{i}e^{r_{i}^{2}t}\text{erfc}(-r_{i}\sqrt{\tilde{t}})\nonumber\\
I_2(r_{i},\tilde{t})&=\frac{2r_{i}^2\sqrt{\tilde{t}}}{\sqrt{\pi}}+(1+2r_{i}^2)r_{i}e^{r_{i}^{2}\tilde{t}}\text{erfc}(-r_{i}\sqrt{\tilde{t}})
\end{align}
At short time $t\ll a^2/D_{0}$ and for the reflecting boundary 
condition,
using the fact $N(\tilde{t})=1-f$, we recover the results of  Eq.~(14)of 
De Swiet and Sen~\cite{deswiet96}:
\begin{equation}
D(\tilde{t})=D_{0}~[1-\frac{4f\sqrt{\tilde{t}}}{3\sqrt{\pi}}+O(\tilde{t})].
\end{equation}
For the fully absorbing boundary condition at short time, we expand
Eq.~(\ref{numerator}) to obtain:
\begin{equation}
\lim\limits_{k \to 0}\frac{\partial G(k,\tilde{t})}{\partial
k^2}=1-\frac{20f}{3\sqrt{\pi}}\tilde{t}^{3/2}+O(\tilde{t}^2).
\end{equation}
This combined with $N(t)$ obtained in Eq.~(\ref{eqn:n}), gives to the
first order of $f$
\begin{equation}
D(t)/D_{0}=1-\frac{2f~\sqrt{\tilde{t}}}{3\sqrt{\pi}}+O(\tilde{t}^2),
\end{equation}
which is consistent with the result in Mitra
\textit{et.~al.}~\cite{mitra2}
Eq.~(A11)~\footnote{Here we take $N(t=0)=N_{0}(1-f)$ instead $N_{0}$ to
be consistent ease the comparison with results in~\cite{mitra2}}.

Much more interesting is the case of fully absorbing boundary
condition at long time $t\gg a^2/D_{0}$. There the numerator of
Eq. (\ref{eqn:num}) reduced to
$\{f(f(6f+29)+19)+5\}\{18f^2(f+1)(f+4)\sqrt{\pi}\}~\tilde{t}^{-3/2}+O(\tilde{t}^{-5/2})$.
Keeping $f$ to first order and combining with
Eq.~(\ref{eqn:r_square}) and the result for $N(\tilde{t})$
Eq.~(\ref{eqn:total}) we obtain the surprising result:
\begin{equation}\label{eqn:long_time}
D(\tilde{t})/D_0=\frac{5}{12f}\tilde{t}^{-1}+O{(\tilde{t}^{-2})},
\end{equation}
that is, within perturbation theory, the long time diffusion
coefficient $D(t)$ of the \emph{dilute} suspension of spheres with
absorbing boundary condition approaches zero as $1/t$. Notice that
$D(\tilde{t})$ is inversely dependent on volume fraction $f$ as is
$N(\tilde{t})$. There is a subtle order of limit: though we are
taking an asymptotic expansion for small $f$, the limit where
Eq.~(\ref{eqn:long_time}) is valid is for a small yet fixed $f$
but $\tilde{t}\to\infty$, i.e. $\tilde{t}\gg 1/f$, where the limit
is well defined in the equation.

Next consider some known results. For diffusion in a straight tube
with fully absorbing wall, the separation of variables can be used
to obtain exact results. The number of particles $N(t)$ will decay
exponentially to zero with time, but $D(t) \rightarrow D_0/3$ as
$t\rightarrow\infty$ because the particles may diffuse along the 
axis of the tube.
Any long tube-like open pore with slowly changing diameter will
also have finite diffusion coefficient. Furthermore a periodic array of
spheres with fully absorbing boundary condition always have finite
diffusion coefficient at long time at any volume fraction. In fact
there is no known connected system,  excluding isolated-pore like
structures (e.g. in the Lifshitz limit\cite {grassberger82}where the
absorbers form a cavity),  with any amount of surface relaxation, which
has a vanishing long time diffusion coefficient. Intuitively a
\emph{dilute} random suspension of spheres is not likely to be an
exception. So the perturbative result of $1/t$ behavior of $D(t)$
may indicate a rapid drop of diffusion coefficient at a certain
range of $t$ but is unlikely to give the correct \emph{true} long
time behavior.

\section{Conclusion}

The $1/t$ time dependence for $D(t)$ given by the perturbative
result for random suspension  is peculiar, as is the $1/t^{3/5}$
behavior given by the heuristic arguments of Grassberger and
Procaccia\cite {grassberger82}. While analytical perturbative
method considered  here give deep insight and give correct result
for short-time limit, we suspect that perturbative results for the
long time behavior of $D(t)$  will break down just as the
perturbative result fails for total number of surviving particles
$N(t)$ at long times. In view of the findings of Fixman
\cite{fixman}, it will be interesting to examine the perturbative
results for $D(t)$ against numerical results with a hope that for
the time-regime of experimental interest, perturbative results will
suffice.
\begin{acknowledgments}
The authors would like to thank  B.~I. Halperin and A.~M. Turner for a 
careful reading of the manuscript. JQ would like to acknowledge the 
people of the NMR group at Schlumberger--Doll Research for their kind
hospitality. This work was also supported in part by the National
Science Foundation grant DMR-99-81283.
\end{acknowledgments}


\begin{thebibliography}{48}
\bibitem{mitra2}
P. P. Mitra, P. N. Sen and L. M. Schwartz, Phys. Rev. B {\bf 47},
8565 (1993). Eq. (30) has severe typographical mistakes which can
be fixed by multipying the right side by $ (\kappa D_0 + \rho )/
(\kappa D_0 - \rho )$ Also, a factor of $S/V_p$ is missing from
the last term in Eq. (29)
\end{thebibliography}
\end{document}